\documentclass[12pt]{article}
\usepackage{psfig,epsfig}
\usepackage{latexsym}
\usepackage{pifont}

\newcommand{\lapprox}{\lower.6ex\hbox{$\; \buildrel<\over\sim \;$}}

\newcommand{\gapprox}{\lower.6ex\hbox{$\; \buildrel>\over\sim \;$}}

\newcommand{\curly}{\lower1.ex\hbox{$\; \stackrel{\textstyle \wr}{\wr} \;$}}

\def\dtheta{\Delta\theta}
\def\Om{\Omega_m}
\def\Ol{\Omega_{\Lambda}}
\def\Ox{\Omega_{x}}

\def\be{\begin{equation}}
\def\ee{\end{equation}}

\def\berr{\begin{eqnarray}}
\def\err{\end{eqnarray}}

\def\apj{{\it Astrophys.~J.}}
\def\aj{{\it Astronom.~J.}}

\def\prd{{\it Phys.~Rev.~D.}}

\def\mnras{{\it Mon.~Not. R.~Astr.~Soc.}}
\def\ijmp{{\it Int.~J.~ Mod. Phys.}}

\def\AnA{{\it Astr. Astrophys.}}

\newcommand{\D}{\Delta}

\renewcommand{\t}{\theta}

\title{ Dark Energy and the Statistical Study of the
Observed Image Separations of the Multiply Imaged Systems in the 
CLASS Statistical Sample }

\author{Abha Dev\footnote{E--mail :
abha@ducos.ernet.in},\,\,
Deepak Jain\,\,\,\footnote{At Deen Dayal Upadhyaya College,
University of Delhi, Delhi 110015} and Shobhit Mahajan \\
	{\em Department of Physics and Astrophysics} \\
        {\em University of Delhi, Delhi-110 007, India} 
	}

\begin{document}

\maketitle


\begin{abstract}

The present day  
observations favour a universe which is flat, 
accelerated and composed of $\sim 1/3$ matter (baryonic + dark) and
$\sim 2/3$ of a negative pressure component, usually referred to as
dark energy or quintessence. The Cosmic Lens All Sky Survey (CLASS), 
the largest radio-selected
galactic mass scale gravitational lens search project to date, has resulted in 
the largest sample suitable for statistical analysis. 
In the work presented here,
we exploit observed image separations of the multiply imaged lensed
radio sources in the sample. We use two different tests:  
(1) image separation distribution function $n(\Delta\theta)$
of the lensed radio sources and (2) ${\dtheta}_{\mathrm{pred}}$ 
vs ${\dtheta}_{\mathrm{obs}}$ as observational tools to constrain the
cosmological parameters $w$ and $\Om$. The results are in concordance
with the bounds imposed by other cosmological tests.

\end{abstract}

\baselineskip= 20pt


\section{Introduction}

A number of observational tests strongly suggest that the universe is
flat and the cosmological energy density includes a component that is
not associated with matter -- baryonic or dark. Results from
distance measurements of type Ia supernovae (SNe Ia) combined with
Cosmic Microwave Background Radiation (CMBR)
observations \cite{garn,perl,efst,sper} and dynamic estimates of the quantity of matter in the
universe seem to indicate that the simple picture provided by standard
cold dark matter scenario is not enough \cite{turner1}. These observations are often
explained by introducing a new hypothetical energy component, usually
referred to as dark energy, with negative pressure. The dark energy is
described by an `equation of state' parameter $w$ ($\equiv
p_{x}/\rho_{x}$).

Since gravitational lensing takes place over cosmological distances,
it can be used to measure the cosmological parameters. Various aspects
of gravitational lensing have been studied and their use to
probe the geometry and constitution of the universe has been
investigated. The frequency of multiply imaged quasars is a sensitive
function of the cosmological constant $\Lambda$ and therefore the observed abundance of lensed
quasars has been used to place upper bounds on $\Lambda$ (Turner 1990 
\cite{elturner1}, Fukugita et al. 1992 \cite{fuku2}, Kochanek 1996
\cite{CSK1}). The relation between the image separation $\dtheta$ and the
source redshift $z_S$ of multiply imaged quasars mostly depends on the
sum of the matter density $\Om$ and vacuum energy density $\Ol$ 
($\Omega_{i}$ is the present density of the $i^{th}$ component of the energy density relative to the
present critical density; $\Omega_{i} \equiv \rho_{i0}/\rho_{c}$), 
and thus serves as a good indicator of the
curvature of the universe. This aspect has been studied by Turner,
Ostriker \& Gott 1984 \cite{TOG}
Gott, Park \& Lee 1989 \cite{GPL}, Fukugita et al. 1992 \cite{fuku2}, Park
\& Gott 1997 \cite{PG}, Williams \cite{williams} (1997) and more recently by
Helbig (1998)\cite{helbig}. Likelihood analysis of the flux
limited samples (established by Kochanek 1996 \cite{CSK1}) is another 
way of obtaining constraints on the
cosmological parameters. This method has been used  extensively in
recent times by many authors to find the permissible range of the
parameters of the cosmological models ( see for example, \cite{CSK1, waga1, chae}). 
Image separation distribution function $n(\D\t)$ has been used to study 
constraints on the cosmological parameters  as well as
on the properties of the galaxies \cite{fuku2,maoz,CY,IJMPD2}. Yet
another aspect of gravitational lensing statistics is the study of the mean
image separation. Gott, Park \& Lee (1989) \cite{GPL} studied the mean image separation of
images in cosmological models with a non-zero cosmological
constant. Lee \& Park
(1994) (hereafter LP) \cite{LP}, working with a sample of five multiply imaged quasars,
used statistical methods to predict the best fit
model. Their choice of models was restricted to six different
combinations of parameters ($\Om$,$\Ol$).  
  
In this work, we exploit the angular image separation of 
lensed quasars in two different ways  to constrain the $\Omega_m$ and $w$: 

\noindent (i) We develop the image separation distribution function 
$n(\dtheta)$ of the multiply imaged lensed radio sources as a 
tool to perform quantitative analysis for finding the constraints on 
the cosmological parameters $w$ and $\Om$. We perform the 
chi-square test to put  
bounds on the parameters. 

\noindent (ii) We calculate the expected value of the angular
separation of a lensed radio source, $\Delta\theta _ {\mathrm{pred}}$ 
by averaging $\Delta\theta$ over the probability distribution function for 
its given redshift $z_S$, flux $S_\nu$ and redshift of the lensing
galaxy $z_L$. To estimate the significance of the difference between the
predicted and the observed $\dtheta$, we once again apply the
chi-square test.

Radio source samples have several advantages over the optical quasar
samples \cite{falco,hel2}. 
Radio-selected sources are immune to extinction in the lens
galaxy. They are also free from the systematic errors arising due to the
fact that the lens galaxy has an apparent brightness 
comparable to that of the lensed
images of the source. The resolution is much smaller than the typical
image separation. The parent catalogues are in the form of large-area
surveys from which unbiased samples can be selected and relatively
easily observed. The Cosmic Lens All Sky Survey (CLASS) has resulted in the 
largest sample of radio sources suitable for statistical analysis 
\cite{CLASS,chae}. We adopt the CLASS statistical
sample for our analyses. The sample consists of 8958 radio sources out of 
which 13 sources are multiply imaged. The sample had been 
selected so as to meet 
well defined observational selection criteria (see Browne et al.(2002)).

The statistical properties of gravitational lensing statistics are
sensitive to the properties of galaxies (lenses) and the sources. 
It is thus important to use reliable
galaxy luminosity functions (LFs) in the analysis of statistical
lensing. We use the luminosity
functions (LFs) for the early-type galaxies (i.e. ellipticals and S0
galaxies) and for the late-type galaxies (i.e. spirals and others) as
provided by the recent large-scale observations of galaxies,  particularly 
the Two Degree Field Galaxy Redshift Survey (2dfGRS) and 
the Sloan Digital Sky Survey (SDSS) \cite{chae}.     

The article is organized as follows: We begin Section 2 by listing the basic
equations of gravitational lensing statistics and the assumptions we
make to derive them. We also briefly describe the adopted source sample and
luminosity function for the galaxies. In Section 3, we develop the
image separation distribution function as a tool to put constraints on
the cosmological parameters. Section 4  describes use of
$\dtheta_{\mathrm{pred}}$ vs $\dtheta_{\mathrm{obs}}$ as a tool
to constrain the parameters $w$ and $\Om$. In Section 5 we discuss our
findings.


\section{\bf  Basic Equations and  Assumptions}
\vskip  0.5cm 

\subsection {Cosmological Model, Distance Formula and Lookback Time} 
 
We consider a spatially flat, homogeneous and isotropic model of
cosmology with non-relativistic matter with energy density $\rho_m$
 and a separately conserved
``dark energy component'' with an equation of state $p_x = w\rho_x$. The
condition for a flat universe becomes $\Omega_m + \Omega_x = 1$ where

\noindent 
 $$ \Omega_m = {8 \;\pi\; G \over 3 H_0^{2}}\rho_{m0}\,\, \mathrm{and}\,\, \Omega_x =
 {8 \;\pi\; G \over 3 H_0^{2}}\rho_{x0}.$$
 $H_0$ is the Hubble
 constant at the present epoch, while $\rho_{m0} \;\;{\rm{and}}
\;\;\rho_{x0} $ are the 
 non-relativistic matter
 density and the dark energy density respectively at the present epoch.

\noindent  The age of the universe at a redshift z is given by
 \begin{eqnarray}
 H_0\; t(z) &= &H_0\;\int_z^{\infty}{dz'\over(1 + z')H(z')}
 \nonumber \\
 & & \nonumber \\
 && =  \int_z^{\infty}{dz'\over(1 + z')\sqrt{\Omega_m(1 + z')^3 +
 \Omega_x(1 + z')^{3(1 + w)}}}. 
 \end{eqnarray}
 \noindent The angular diameter distance between objects at redshifts 
$z_1$ and $z_2$ is  
 \be
 d_{A}(z_1,z_2) = {c {H_{0}^{-1}}\over (1 + z_2)}\int_{z_1}^{z_2}
 {dz\over \sqrt{\Omega_m(1 + z)^3 + \Omega_x(1 + z)^{3(1 + w)}}}.
 \ee

We adopt the notation $D_{LS} = d_{A}(z_L,z_S)$.

\vskip 0.5cm
\subsection{Galaxy Luminosity Function }

\noindent The number density of galaxies with luminosities lying
between $L$ and $ L + dL$ is assumed to be described by the Schechter
function of the form \cite{sch}

\be
 \Phi(L, z = 0)dL \,=\,
 \phi_\ast\, \left({L \over L_\ast}\right )^\alpha \,\,\exp\left (-
 {L\over  L_\ast}\right ) \,{dL\over L_\ast},
 \ee
\noindent where $\phi_\ast, \alpha,$ and $L_\ast$ are the
characteristic number density, faint end slope and characteristic
luminosity respectively at the present epoch. We further assume that the total
comoving number density of galaxies is conserved.
Thus the comoving number density of galaxies at redshift $z$ is
given by $n_{L}(z) = n_{0} (1+z)^3$ where $n_0$ is the present number
density of galaxies. 

\noindent The present day comoving number density of galaxies can be
calculated as 

\be
 n_0 = \int_{0}^{\infty}\Phi(L)dL.
\label{n0}
 \ee

The two large-scale galaxy surveys: 2dFGRS and SDSS, have
produced converging results on the total LF. The surveys determined  
the Schechter parameters for galaxies (all types) at $z \le 0.2$.       
Chae (2002) \cite{chae} has worked extensively on the information 
provided by the recent galaxy surveys to extract the local type-specific
LFs. As the information on the total LF is abundant and the results
are convergent, the total LF can be decomposed into an early-type
LF and a late-type LF. The inference of the type-specific LFs from the
total LF relies on an independent knowledge of relative type-specific
LFs. We use the normalization corrected 2dFGRS lens and Schechter
parameters as evaluated by Chae \cite{chae}. The 2dFGRS survey (Folkes et
al. (1999) \cite{folkes}) employs the ``principal component
analysis'' of galaxy spectra, which essentially maximally quantify
differences between the spectra of galaxies, to study LFs per type.     
Table 1 lists the lens and the Schechter parameters used in this work. We
have adopted the parameters per morphological type as given by Chae
(2002)\cite{chae}.

For the ellipticals and lenticulars, 
characteristic velocity dispersion $v_\ast$ 
and the characteristic luminosity $L_\ast$ follow the Faber-Jackson relation
($L_\ast \propto v_\ast^{4.0}$). For spirals,
 $v_\ast$ and $L_\ast$ are related by
the Tully-Fisher relation ($L_\ast \propto v_\ast^{2.6}$).
We express this as
\be
\frac{L}{L_\ast} = (\frac{v}{v_\ast})^\gamma
\label{FJ}
\ee

\noindent with $\gamma = 4.0 $ for the early-type galaxies and $\gamma =
2.6$ for the late-type.

\begin{table}
\begin{center}
\begin{tabular}{|l|lllll|}\hline\hline
Galaxy & $ \alpha$ & $ \gamma$ & $ v^{*} \mathrm{(km /s)}$ & $\phi^{*}
(\mathrm{Mpc^{-3}})$ & $ F^{*}$ \\
Type  &   & & &  & \\\hline
 \hline
 E/S0 & $-0.74$ & $4.0$ &$185$ &$0.82 \times 10^{-2}$& $0.014$   \\
 S  &  $-1.0$ & $2.6$ &$134$ & $1.2 \times 10^{-2}$& $0.008$ \\
\hline
\end{tabular}
\caption{Lens and Schechter parameters for galaxies 
(for details see \cite{chae}).} 
\end{center}
\end{table}

\vskip 0.5cm
\subsection {Lensing Statistics}

We consider the singular isothermal sphere (SIS) model for the
lens mass distribution because of
simplicity. SIS is a good approximation to the real mass distribution in
galaxies \cite{TOG,maoz}. The cross-section for lensing events for the SIS
lens  is 

\be
\sigma = 16\,\pi^{3}\,({v \over c})^{4}(\frac{D_{OL}D_{LS}}{D_{OS}})^{2}.
\label{s1}
\ee

For the calculations presented here we ignore evolution of number density
of galaxies and assume that the comoving number density is conserved.
Under these assumptions, the differential probability of a lensing
event by a galaxy with velocity dispersion $v$ at redshift $z_{L}$ is

\be
d\tau = n_{0}(1+z_{L})^3\,\sigma \,\frac{cdt}{dz_{L}} dz_{L}. 
\label{d1}
\ee

\noindent The quantity $cdt/dz_{L}$ in Friedman Robertson Walker (FRW)
geometry for the 
adopted cosmological models is 

\be
\frac{cdt}{dz_{L}} = \frac{a_{0}}{(1 + z_{L})}
\,\frac{1}{\sqrt{\Om(1 + z)^{3} + \Ox (1 + z)^{3(1 + w)}}}
\ee

\noindent where $a_{0}$ is the scale factor at the present epoch.

Substituting for $\sigma$ from equation (\ref{s1}), we get
 
\be 
d\tau = n_0 (1+z_{L})^{3}\,{16 \pi^{3}\over {c H_{0}^{3}}}\,
v^{4}\,({D_{OL}D_{LS}\over a_{0} 
D_{OS}})^{2}\,{1\over a_{0}}\,{cdt\over dz_{L}}\,dz_{L}. 
\label{r1}
\ee

Using equations (\ref{n0}), (\ref{FJ}) and (\ref{r1}),
 the differential optical depth of lensing in traversing
 $dz_{L}$ with angular separation between $\phi$ and $\phi + d\phi$
 is  \cite{fuku2}:

 \begin{eqnarray}
 {\frac{ d^{2} \tau }{ dz_{L}d\phi}}\,{d\phi}\,{dz_{L}}
 &=&{{\gamma} \over
 2 \phi } \left [{D_{OS}\over{D_{LS}}} \phi
 \right ]^{{\frac{\gamma}{2}}(\alpha + 1+ {\frac{4}{\gamma}})}
 \exp\left[-
 \,\left({D_{OS}\over{D_{LS}}} \phi\right)^{\frac{\gamma}{2}} \right
 ]{c\,dt\over dz_{L}} \nonumber \\
 & & \nonumber \\
 &&\times \, F^{*}\,{(1 + z_{L})^{3} \over \Gamma\left(\alpha +
 {4\over\gamma} +1\right)}\left[\,\left ({D_{OL}D_{LS}\over a_0
 D_{OS}}\right)^{2}\,{1\over a_0}\right ]\,{d\phi}\,{dz_{L}},
 \label{d2}
 \end{eqnarray}

\noindent where $F^*$ is defined as

\be
F^{*} ={16\pi^{3}\over{c
H_{0}^{3}}}\,\phi_\ast\, v_\ast^{4}\Gamma\left(\alpha +
{4\over\gamma}+1\right).
\ee

The value of $F^*$ for the adopted lens and Schechter parameters for
the early and the late-type galaxies is given in Table 1.

\vskip 0.5cm
\subsection { The CLASS Statistical Sample}

 The sample consists of
8958 radio sources out of which 13 sources are multiply imaged. While
using $n(\dtheta)$ as a tool to constrain the cosmological parameters,
we work only with those multiply imaged sources whose image-splittings
are known (or likely) to be caused by single galaxies. There are 9
such radio sources: 0218+357, 0445+123, 0631+519, 0712+472, 0850+054,
1152+199, 1422+231, 1933+503, 2319+051. (See Chae (2002) \cite{chae}
for the complete and detailed list of the lensed sources.) 
We thus work with a total of 8954
radio sources. The sources probed by CLASS at $\nu = 5$ GHz 
are well represented by
power-law differential number-flux density relation: $\left |dN/dS
\right| \propto (S/S^{0})^{\eta}$ with $\eta = 2.07 \pm 0.02$ ($1.97 \pm
0.14$) for $S \geq S^{0}$ ($ \leq S^{0}$) where $S^{0} = 30$ mJy
(McKean et al.(2002)). 
The redshift distribution of unlensed sources in the sample can be
adequately described by a Gaussian model
with a mean redshift $z = 1.27$ and dispersion of $0.95$ \cite{chae}.
Guided by the above information about (i) the number-flux density relation 
and (ii) the distribution of unlensed sources in redshift, 
we simulate the unlensed radio sources (8945 in number) of the
CLASS statistical sample using the Monte-Carlo technique (rejection method).

\vskip 0.5cm
\subsection{ Magnification Bias}
 
\noindent  Gravitational lensing causes a magnification of
images and  this transfers the  lensed sources to higher flux
density bins. In other words, the lensed quasars are over-represented in a
flux-limited sample. The magnification bias  ${\mathcal B}(z_S,S_\nu)$ increases
the lensing probability significantly in a bin of total flux density
($S_\nu$) by a factor

\be
\mathcal{B}(z_S,S_\nu) = \left |\frac{dN_{z_S}(>S_\nu)}{dS_\nu}
\right|^{-1}\int_{\mu_{min}}^{\mu_{max}} \left |\frac{dN_{z_S}
(>S_\nu/\mu)}{dS_\nu}\,p(\mu)\right |{1\over\mu}
\,d\mu.
\label{B2}
\ee

\noindent Here  $N_{z_{S}}(> S_\nu)$ is the intrinsic flux density
relation for the source population at redshift $z_{S}$. $N_{z_{S}}(> S_\nu)$
 gives the number of sources at redshift $z_{S}$ having flux greater than $S_\nu$. 
For the SIS model, the magnification probability distribution is
 $p(\mu) = 8/{\mu}^{3}$. The minimum and maximum total magnifications $\mu_{min}$ and
$\mu_{max}$ in equation (\ref{B2}) depend on the observational
characteristics as well as the lens model. For the SIS model, the
minimum total magnification is $\mu_{min} = 2$ and the maximum total 
magnification is $\mu_{max} = \infty$. For the CLASS statistical
sample, one of selection criteria for the double-imaged systems is that
the ratio of the flux densities of the fainter to the brighter images
${\mathcal R}_{min}$ is $ \ge 0.1$. Given such an observational limit,
the minimum total magnification for double imaging the adopted  model
of the lens is \cite{chae}: 

\be
\mu_{min} = 2 \frac{1+\mathcal{R}_{min}}{1-\mathcal{R}_{min}}.
\ee  

The magnification bias $\mathcal B$ depends on the differential
number-flux density relation \\
$\left|dN_{z_{S}}(> S_{\nu})/dS_{\nu}
\right|$. The differential number-flux relation needs to be known as a
function of the source redshift. At present redshifts of only few
CLASS sources are known. We, therefore, ignore redshift 
dependence of the differential number-flux density relation. Following
Chae (2002) \cite{chae}, we further ignore the dependences of the differential
number-flux density relation on the spectral index of the source.

\vskip 1.0cm
\section{$n(\dtheta)$ As a Probe}

The normalized image angular separation distribution for  a source at
 $z_{S}$ is obtained by integrating $\frac{d^{2}\tau}{dz_{L}\,d\phi}$
over $z_{L}$:

 \be
 {d{\mathcal{P}}\over d\phi}\, =\, {1\over\tau(z_S)}\int_{0}^{z_{s}}\,{\frac{
 d^{2}
 \tau }{dz_{L}d\phi}} {dz_{L}}.
 \ee

The corrected image separation distribution function
for a single source at redshift $z_{S}$ is given by  \cite{CSK1,CY}

 \begin{eqnarray}
 P'(\Delta\theta)\,& =& \, \mathcal{B}\,{{{\gamma}} \over
 2\, \dtheta} \int_{0}^{z_S}\left [{D_{0S}\over{D_{LS}}} \phi
 \right ]^{{\frac{\gamma}{2}}(\alpha + 1+ {\frac{4}{\gamma}})}
 \exp\left[-
 \,\left({D_{0S}\over{D_{LS}}} \phi\right)^{\frac{\gamma}{2}} \right
 ]\nonumber \\
 & & \nonumber \\
 &&\times \, F^{*}\,{cdt\over dz_{L}} {(1 + z_{L})^{3} \over
\Gamma\left(\alpha + 
 {4\over\gamma} +1\right)}\left[\,\left ({D_{OL}D_{LS}\over a_0
  D_{OS}}\right)^{2}\,{1\over a_0}\right ]\,\,{dz_{L}}.
\label{dist}
 \end{eqnarray}

\noindent Similarly the corrected lensing probability for a given
source at redshift $z$ is given  by \cite{CSK1,CY}

\be
P' = \,\int {d{\mathcal{P}}\over d\phi} \mathcal{B}\; d\phi.
\label{prob}
\ee

\noindent Here $\phi$ and ${ \Delta\theta}$ are related through
 $\phi = {\frac{\Delta\theta}{8 \pi (v^{*}/c)^2}}$.

The expected number of lensed radio sources is $n_{\rm L} =
\sum P'_{i}$, where $P'_{i}$ is the lensing probability of 
the $i^{th}$ source and the sum
is over the entire adopted sample. Similarly, 
the image-separation distribution 
function for the adopted sample is $n(\Delta\theta) 
= \sum P'_{i}( \Delta\theta)$. 
The summation is over all the radio sources in the sample.

For the study of image separation distribution function, we 
neglect the contribution of spirals as lenses because their
velocity dispersion is small as compared to ellipticals (Table 1).

 For a given $\Om$ and $w$, we
 compare the predicted number of multiply imaged sources in 
 $\dtheta$  bins with the observed number
 of lensed radio sources in the corresponding bins. 
As one of the selection criteria for
 the multiply-imaged systems is that the compact radio core images have 
separations $\geq 0.3$ arcseconds, we compare the image separation
 distributions for $\dtheta \geq 0.3$ arcseconds. Table 2 gives 
the observed image separation distribution of the 
multiply imaged radio sources. 
We perform the chi-square test to find bounds on $w$ and
 $\Om$. We search for the combination ($\Om,w$) for which 
the value of chi-square
becomes minimum. We define the chi-square as:

\be
\chi^2 = \sum_{k = 1}^{6}\frac{ (n_{\mathrm{pred}\, k} -
 n_{\mathrm{obs}\, k})^2}{ n_{\mathrm{obs}\, k}}.
\ee

\noindent Here $n_{\mathrm{pred}\, k}$ is the predicted number of 
lensed sources with
image separation in the interval corresponding to the $k^{th}$ bin
and is given by the area under the curve $n(\dtheta)$ for the bin
under consideration. $n_{\mathrm{obs}\, k}$ is the observed number of multiply
imaged CLASS sources in the $k^{th}$ bin. We work with the following ranges
of the parameters: $-1 \le w \le 0$ and $0 \le \Om \le 1$ and perform a grid
search in the parametric space to find the best fit model.
For the two parameter fit, the 68\% Confidence Level (CL) 
(90\% CL) corresponds to $\Delta\chi^2 = 2.3$
($4.61$).

\begin{table}
\begin{center}
\begin{tabular}{|l|llllll|}\hline\hline
$\dtheta$ intervals: & $0.3-0.8$ & $0.8-1.3$ & $1.3-1.8$ & $1.8-2.3$ & $2.3-2.8$
  & $2.8-3.3$ \\   
(arcseconds) &     &     &   &    &    &  \\ \hline
Number of & $2$ & $4$ &$3$& $0$ & $0$ & $0$   \\
lensed sources    &     &     &   &     &     & \\
\hline
\end{tabular}
\caption{Observed distribution of the 9 multiply-imaged sources in
 the CLASS statistical sample.} 
\end{center}
\end{table}

\vskip 1.0cm
\section{${\dtheta}_{\mathrm{pred}}$ vs ${\dtheta}_{\mathrm{obs}}$ As
  an Observational Tool} 

With the lensing galaxy at redshift $z_L$, the mean image separation of
images of a multiply imaged source at
redshift $z_S$ is given by \cite{LP}:

\be
\dtheta_{\mathrm{pred}} = \langle \dtheta \rangle = \frac{\sum 
\int\frac{d^{2}\tau}{dz_{L}\,d\dtheta}\,\dtheta
d\dtheta}{\sum
\int\frac{d^{2}\tau}{dz_{L}\,d\dtheta}\,d\dtheta},
\label{dt1}
\ee

\noindent where the summation is over the galaxy types, E/S0 and S.

The expected dispersion of separation around the mean value is

\be
\sigma_{\dtheta} = \sqrt{\langle \dtheta^{2}\rangle - \langle \dtheta
\rangle ^{2}}
\ee

\noindent with

\be
\langle \dtheta^{2} \rangle = \frac{\sum 
\int\frac{d^{2}\tau}{dz_{L}\,d\dtheta}\,\dtheta^2
d\dtheta}{\sum
\int\frac{d^{2}\tau}{dz_{L}\,d\dtheta}\,d\dtheta}.
\label{dt2}
\ee

We again focus our attention on
the multiply imaged radio sources whose image-splittings are known (or
likely) to be caused by single galaxies. Further, we consider only
those lensed sources for which the source redshift $z_S$, 
redshift of the lensing galaxy $z_L$ and observed image separation
$\dtheta_{obs}$ are known. There are 6 such sources (Table 3).

\begin{table}
\begin{center}
\begin{tabular}{|l|lllll|}\hline\hline
Source & $ \dtheta_{\mathrm{obs}}$ & $ z_{S}$ & $ z_{L}$& $\dtheta_{\mathrm{pred}}$ 
&$\dtheta_{\mathrm{pred}}$ \\ 
       &  (arcseconds)   &   &   & (arcseconds) & (arcseconds) \\  
       &             &   &   & no sel.  & sel.        \\\hline
0218+357 & $0.334$ &$0.96$ & $0.68$& $0.50$  & $0.62$\\
0712+472 & $1.27$  &$1.34$ & $0.41$ & $1.19$ & $1.26$\\
1152+199 & $1.56$  &$1.019$ & $0.439$& $0.98$& $1.06$ \\
1422+231 & $1.28$  &$3.62$ & $0.34$ & $1.56$ & $1.62$\\
1933+503 & $1.17$  &$2.62$ & $0.755$& $1.22$ & $1.29$\\
2319+051 & $1.36$  &$2.0$&$0.624$& $1.18$ & $1.26$\\
\hline
\end{tabular}
\caption{We take $z_S = 2$
for the source (2319+051) of unknown redshift, which is the mean redshift of
the whole sample \cite{chae}. The fourth column gives the corresponding
predicted image separations for the best fit model: $\Om = 0$ 
\& $w = -1$ when selection
effects are not considered. The fifth column gives the 
predicted values of $\dtheta$
when selection effects are taken into account.} 
\end{center}
\end{table}

We work with the following ranges of the parameters: $-1 \le w \le 0$
and $0 \le \Om \le 1$.
To find the best fit model we apply the chi-square test. The 
chi-square is defined as 

\be
\chi^{2} =
\sum_{i=1}^{N} \left [ \frac{\dtheta_{\mathrm{obs}\,i}-
\dtheta_{\mathrm{pred}\,i}}
{\sigma_{\dtheta \, i}} \right ]^{2}.
\ee

The sum is over all the data points ($N = 6$). The chi-square per degree
of freedom for a two parameter fit is defined as 
$\chi^{2}_{\nu} = \chi^{2}/(N-2)$. 
We perform a grid search in the parametric space to find the best fit
model, the one for which value of $\chi^{2}$ is minimum.

It is difficult to observe gravitational lens systems of multiple
images having very small angular separations. As mentioned above (section 3), 
one of the selection criteria for
the multiply-imaged system is that the compact radio core images have 
separations $\geq 0.3$ arcseconds. 
We, therefore, also work out the constraints tuned to this selection effect.  

\section{Results and Discussions}

Image separation distribution function $n(\D\t)$ has been used to study 
constraints on the cosmological parameters  as well as
on the properties of the galaxies \cite{fuku2,maoz,CY,IJMPD2}.  
$n(\dtheta)$ is sensitive to
the cosmological parameters via the angular diameter distance formula.
Earlier work on $n(\dtheta)$ does not employ any
statistical methods to put bounds on the parameters. As explained in
section 3 , we use the chi-square test to find the best fit model as well
as the permissible range of $w$ and the matter density $\Om$. 
It is interesting to note that $\chi^{2}$ attains a minimum value of $\sim 3.1893$
(implying $\chi^{2}_{\nu \mathrm{min}} = 0.8$) for the combinations 
($\Om$,$w$), as indicated in Fig. 1. For these
combinations the total number of lensed sources for the adopted
sample, which is defined by the total area under the curve 
$n(\dtheta)$ vs $\dtheta$ (with $\dtheta \leq 3.3$), works out to
be $\sim 10.6$. Fig. 1 also shows the contours of  68\% and 90\% CL. The analysis
provides us with the following constraints: $w \leq -0.13$ and $\Om \leq
 0.74$ at 68 \% CL and $w \leq -0.05$ and $\Om \leq 0.91 $ at 90\% CL.

Fig. 2 compares the observed distribution of the lensed sources in $\dtheta$ 
with the predicted distribution for
$\Om = 0.3$ and $w = -0.585$ (one of the points in the 
parametric space for which
$\chi^2$ is minimum).

The cosmological parameters determine the number of the predicted 
lensed sources in a bin and hence the height of the predicted histogram. 
There is, however, no shift in
the peak of the predicted histogram with the change in the values ($\Om$,$w$). On
the other hand, the location of the peak of the $n(\dtheta)$ curve
strongly depends on the model of galaxy evolution \cite{d,abha}. The
tool developed in this piece of work can prove useful to 
study models of galaxy evolution. The image separation distribution 
function also depends on the mass profile of the lensing 
galaxies \cite{HM}. The above test can be used to suggest the  
preferred lens model or to constrain the parameters of the model under 
consideration.

In this article we also exploit the mean image separation as a tool to
constrain the parameters $\Om$ and $w$. The mean image separation 
of the multiply
lensed quasar is sensitive to the parameters via the 
differential lensing probability.  

It is interesting to note that the value of chi-square becomes minimum for 
$\Om = 0.0$ and $w = -1$, with $\chi^{2}_{\nu \,\mathrm{min}} 
= 0.412 $ ($0.66$) for the
combination with no selection effects (with selection effects).  
This result is in agreement with
the observations of Lee \& Park \cite{LP}. They consider a flat universe with a
non-zero cosmological constant and a sample of five gravitational
lens systems. They report that among the six different models that
they have considered, the ones with large $\Ol$ are preferred.
 
Fig. 3 shows the contour of $68 \% $ CL in the parametric space when 
`no selection effect' is considered. 
The best-fit model for which the chi-square attains the minimum value is $\Om
=0.0$ and $w = -1$ . At 68 \% CL, we get $\Om \le 0.5$ and $w \le
-0.24$. The entire parametric space is allowed at 90\% CL. 

Fig. 4 shows the results obtained when we incorporate the selection effects.
The best fit model is again $\Om =0.0$ and $w = -1$. The constraints obtained 
are weaker when we take selection effects into account. 
The bounds are: $\Om \leq 0.61$
and $w \leq -0.16$ at 68 \% CL. The entire parametric space 
is allowed at 90 \% CL.

The fourth and the fifth columns of Table 3 give the values 
of image separations as predicted by the best fit model. 
Like LP, we also observe that $\dtheta_{\mathrm{pred}}$ is larger 
when selection effects are taken into account.

Table \ref{T4} lists the constraints obtained on $w$ and $\Om$ from 
various aspects of gravitational lensing statistics along with those
obtained from other observational tests. 

The observed statistical properties of gravitational lensing in a
sample are the total rate of lensing, the lens redshifts, the source
redshifts and the image multiplicity. In this body of work, we 
explored the use of image separation as an observational tool. 
The bounds obtained in this article are concordant with those obtained 
from various other observational tests (see Table 4). 
Nevertheless, the constraints derived are 
exposed to the following possible sources of errors. We first point out those 
sources of errors which affect both the analyses.

(i) Like other aspects of gravitational lensing statistics, the 
observational tests presented here are sensitive to 
the lens and the Schechter parameters.
At present, there is no consensus amongst the observationally derived
results for early-type LF \cite{chae}. The main difficulty in the
determination of the early-type LF lies in  identifying a large number
of galaxies by morphological types. In this piece of work we have used
the current estimates of luminosity functions of galaxies per
morphological type. The present derived constraints on the
cosmological parameters are susceptible to systematic errors arising
from the uncertainties in the early-type LF.

(ii) We adopt the SIS model for the lensing galaxies which is an over
simplification. 

(iii) In this work we work with the assumption that 
the early-type galaxies do not evolve with time. The presence of 
number evolution of galaxies in addition to pure luminosity evolution
decreases the optical depth which makes the constraints on $w$ weaker
\cite{DJ1,abha}. The inclusion of the evolution factor also affects
the image separation distribution \cite{abha}.
 
The results of `$n(\dtheta)$ test' are further susceptible to the following
sources of error:

\noindent $\bullet$ At present, redshifts of only a 
few CLASS sources are known and this
affects the analysis in several ways. For instance,
we are led to assume that the number-flux density 
relation of the radio sources in the CLASS statistical sample is
independent of the  source redshift $z_{S}$. This assumption adds to
the possibility of an error.
Further as Chae (2002) \cite{chae} has shown that the derived value
of $\Om$ (in a flat universe) is sensitive to the mean redshift. Thus
there can be an error arising from the current uncertainty in the mean redshift.

\noindent $\bullet$ While using $n(\dtheta)$ as a probe we have ignored the use
of spirals as lenses. This may be another possible source of error.

While using ${\dtheta}_{\mathrm{pred}}$ vs ${\dtheta}_{\mathrm{obs}}$ 
as a tool, we are led to work with a sample size of 6 multiply
imaged radio sources. Availability  of  more data on the redshifts 
of the lenses and the radio sources as
well as the lens parameters  will help to use the tools described
above in a more powerful way.

\begin{table}
\begin{center}
\begin{tabular}{|l|l|l|r|}\hline\hline
Method             &  Reference      & $w$    &   $\Om$  \\ \hline
Large scale structure&  Garnavich et al.\cite{garn}, & $w \leq -0.6$
& $0.3 \leq \Omega_m \leq 0.4$\\
+CMBR+SNe Ia data   &Perlmutter et al.\cite{perl}    & at 95 \% CL&  at 95 \% CL \\
                   & Efstathiou \cite{efst}) &        &             \\
                   &                 &        &             \\
    CMB            & Spergel \cite{sper}&$w \leq -0.78$ & $\Om = 0.27 \pm 0.04$ \\
                   &                           & at 95 \% CL  & at 95 \% CL \\
                   &                           &              &           \\                    
Old High  &Lima and Alcaniz\cite{alcaniz1}& $w \leq -0.2$ & $\Om =0.3$ \\
Redshift Galaxies     &                               &      &    \\
& & & \\
Angular Size- & Lima and Alcaniz \cite{alcaniz2}& $ w \leq -0.2$  & $\Om \leq 0.62$ \\
 Redshift Data & (For $l\simeq 20 h^{-1} Mpc$)               & at 68 \% CL      & at 68 \% CL \\       
& & &\\
X-ray Clusters & Schuecker et al. \cite{Schu} & $ w = -0.95^{+0.30}_{-0.35}$&
$\Om = 0.29^{+0.80}_{-0.12}$ \\
of Galaxies & & at $ 1 \sigma$ & at $ 1 \sigma$ \\
& & & \\
Lensing Statistics &                 &        &           \\
(i) Likelihood     & Waga and Miceli\cite{waga1}& $w \leq -0.71$ & $0.24\leq \Omega_m \leq 0.38$\\
\hspace{0.5cm} Analysis  & (optical sample) &       at 68\% CL &   at
68\% CL \\
                   &               &        &    \\
                   & Chae \cite{chae}       & $w < {-0.55}^{+0.18}_{-0.11}$&$\Omega_m =
{0.31}^{+0.27}_{-0.14}$\\              
                   & (CLASS sample)         & at 68\% CL & at 68\%
CL\\
                   &             &   &  \\
(ii)$n(\dtheta)$   & {\bf This paper}  & $w \leq -0.13$  & $\Om \leq 0.74$  \\
                   & (CLASS sample) & at 68 \% CL & at 68 \% CL \\
(iii)${\dtheta}_{\mathrm{pred}}$ Vs ${\dtheta}_{\mathrm{obs}}$ & {\bf
  This Paper} &  $w \le-0.24$ & $\Om \le 0.5$\\
              & (CLASS sample)& at 68 \% CL & at 68 \% CL  \\
                   &                & & \\                     
\hline 
\end{tabular}
\caption{Constraints on $w$ \& $\Om$ from various cosmological tests.}
\label{T4}
\end{center}
\end{table}

\begin{section}*{Acknowledgement}

We are thankful to Kyu -Hyun Chae, Phillip Helbig and Amber Habib
for useful discussions. We also wish to thank Chris Kochanek, 
Lindsay King and Zong-Hong Zhu for their useful comments on the earlier 
version of the paper.

\end{section}


\eject

\begin{figure}
\vskip 19 truecm
\includegraphics{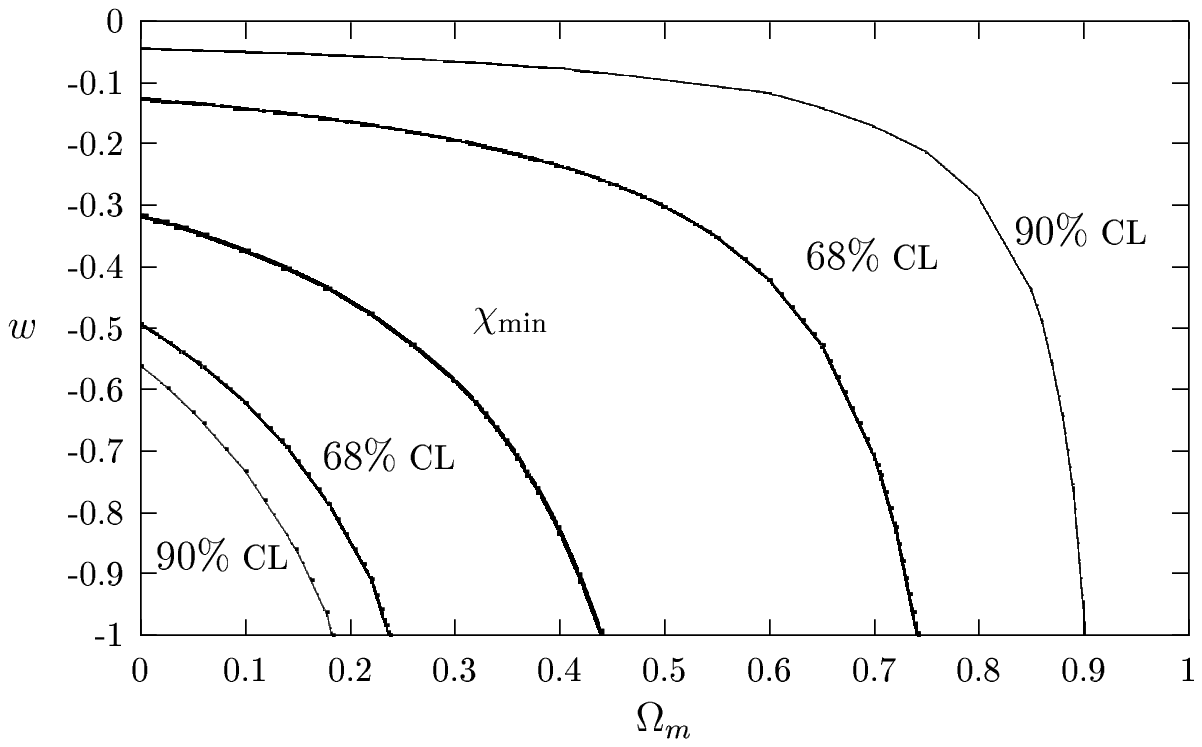}
\caption{The contours showing permissible
region of the parametric space within 68\% CL and 90\% CL. 
The contour labeled $\chi_{\mathrm{min}}$ corresponds to
those combinations of $\Om$ and $w$ for which the value of chi-square is minimum.}
\end{figure}
\vfill
\eject

\begin{figure}
\vskip 19 truecm
\includegraphics{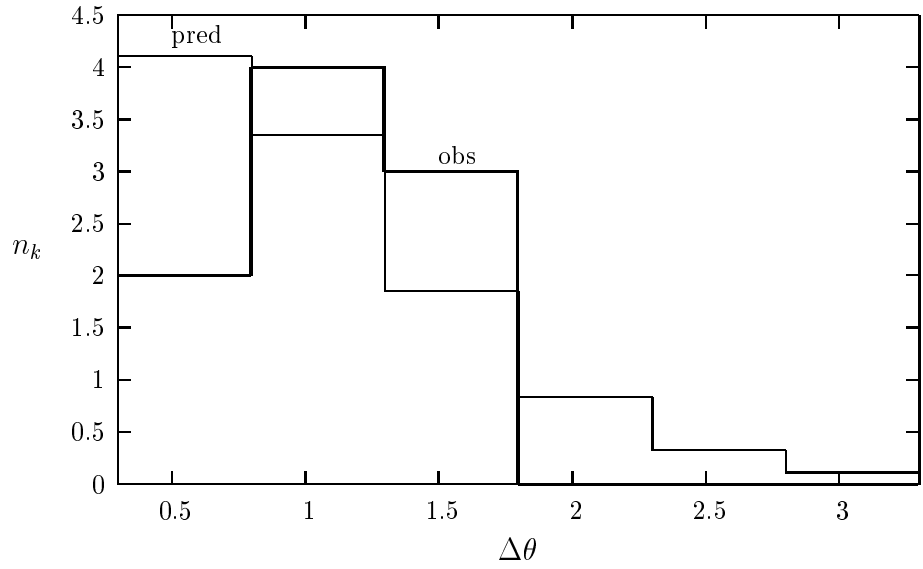}
\caption{The observed distribution of lensed radio sources in $\dtheta$ vs the
predicted distribution for $\Om = 0.3$ and $w = -0.585$.}
\end{figure}
\vfill
\eject

\begin{figure}
\vskip 19 truecm
\includegraphics{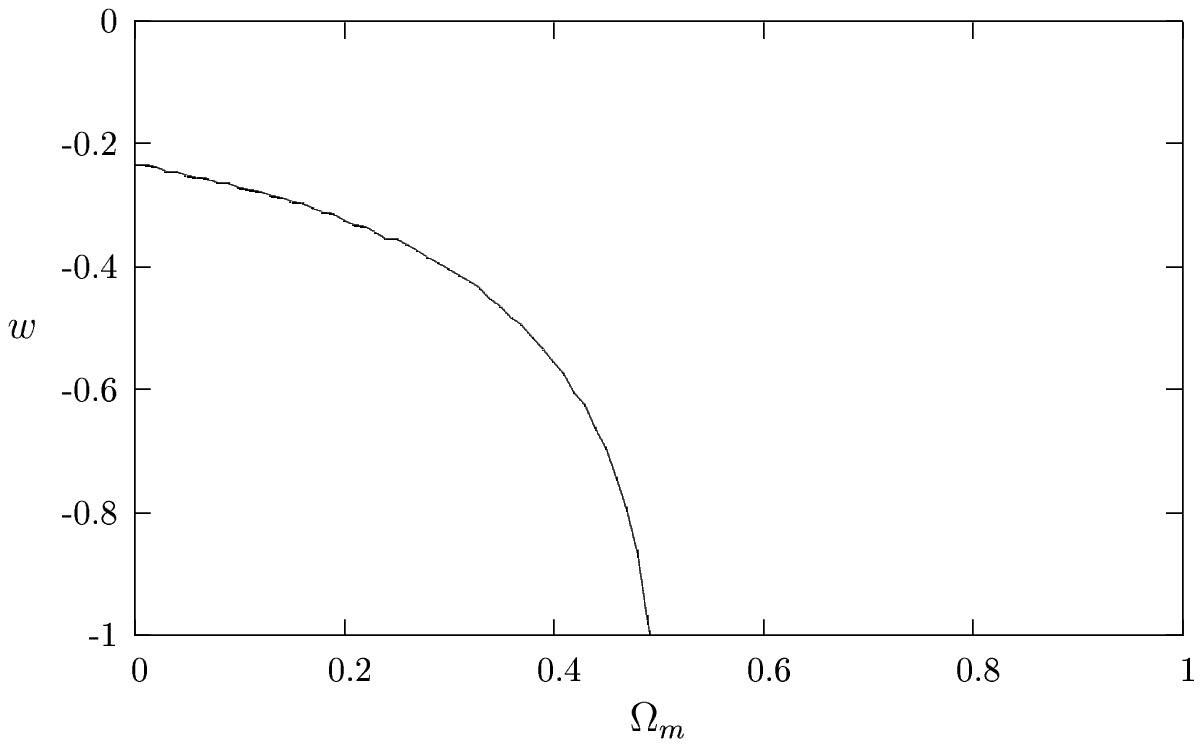}
\caption{ Results of ${\dtheta}_{\mathrm{pred}}$ Vs ${\dtheta}_{\mathrm{obs}}$
test.
The graph shows the permissible region of $\Om$-$w$ plane when 
selection effect is not considered. The contour corresponds to 68\% CL, the
best fit model being: $\Om =0$ with $w = -1$. The contour corresponding
to 90\% CL lies outside the range chosen for $(\Om,w)$}
\end{figure}
\vfill
\eject

\begin{figure}
\vskip 19 truecm
\includegraphics{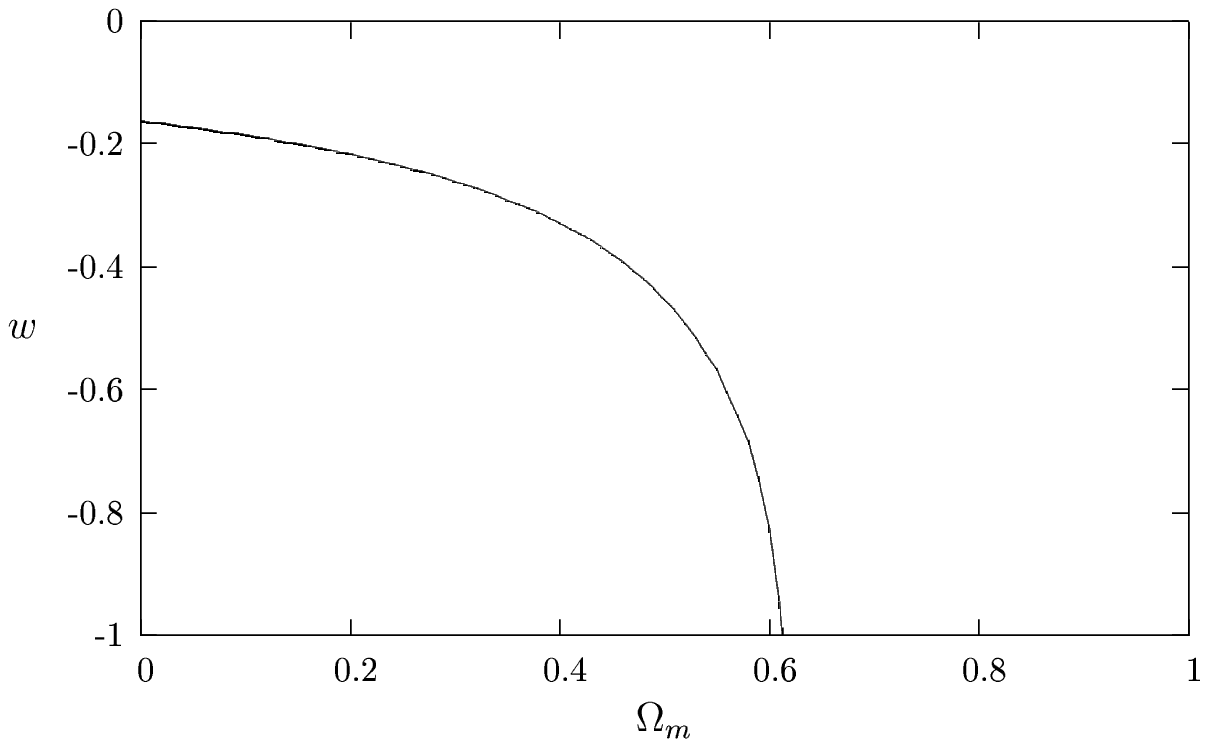}
\caption{ Results of ${\dtheta}_{\mathrm{pred}}$ Vs ${\dtheta}_{\mathrm{obs}}$
test. The graph shows the permissible region of $\Om$-$w$ plane when selection effect 
is taken into account. 
The contour corresponds to 68\% CL, the
best fit model being: $\Om =0$ with $w = -1$. The contours corresponding
to 90\% CL lies outside the range chosen for $(\Om,w)$}
\end{figure}
\vfill
\eject

\end{document}